\documentclass[preprintnumbers,twocolumn,superscriptaddress,aps,nofootinbib]{revtex4}
\usepackage{amssymb}
\usepackage[centertags]{amsmath}
\usepackage{txfonts}
\usepackage{epsfig}
\usepackage{bm}
\usepackage{color}
\usepackage{graphicx,graphics}
\usepackage{multirow}
\usepackage{float}
\usepackage[pdfstartview=FitH]{hyperref}
\hypersetup{colorlinks=true,citecolor=blue,linkcolor=blue,filecolor=black,urlcolor=blue}
\allowdisplaybreaks[2]

\begin{document}
\title{Nuclear suppression of azimuthal asymmetries in semi-inclusive deep inelastic scattering off polarized targets}

\author{Yu-kun Song}
 \affiliation{Department of Modern Physics, University of Science and Technology of China, Hefei, Anhui 230026, China}
\author{Zuo-tang Liang}
 \affiliation{Key Laboratory of Particle Physics and Particle Irradiation (MOE) \& School of Physics, Shandong University, Jinan 250100, China}
\author{Xin-Nian Wang}
 \affiliation{Key Laboratory of Quark \& Lepton Physics (MOE) and Institute of Particle Physics, Central China Normal University, Wuhan, 430079, China}
 \affiliation{Nuclear Science Division, MS 70R0319, Lawrence Berkeley National Laboratory, Berkeley, California 94720}

\date{\today}


\begin{abstract}
We extend the study of nuclear dependence of the transverse momentum dependent
parton distribution functions and azimuthal asymmetries to semi-inclusive deep inelastic scattering (SIDIS)
off polarized nuclear targets. We show that azimuthal asymmetries are suppressed for SIDIS off a polarized nuclear target
relative to that off a polarized nucleon due to multiple scattering inside the nucleus. Using the value of transport parameter
inside large nuclei extracted from jet quenching analyses in SIDIS off nuclear targets, we also present
a numerical estimate of the nuclear suppression of the azimuthal asymmetry that might be useful to guide the future
experimental studies of SIDIS off polarized nuclear targets.
\end{abstract}


\maketitle

{\it Introduction}.
Azimuthal asymmetries in semi-inclusive deep-inelastic lepton-nucleon scattering (SIDIS)
are sensitive probes of the transverse momentum dependent (TMD)
parton distributions and/or correlation functions in a nucleon. In SIDIS off nuclear targets,
multiple parton scattering off different nucleons inside the nucleus can lead to many interesting
phenomena such as suppression of leading hadrons \cite{0005044,0102230},
transverse momentum broadening \cite{9804234,Kang:2013raa}
and nuclear modification of the azimuthal asymmetries \cite{Gao:2010mj,Song:2013sja}.
In a recent publication \cite{Song:2013sja}, we calculated azimuthal asymmetries
in semi-inclusive DIS  process $e^-+N\rm{\ or\ } A\to e^-+q+X$ for both polarized and unpolarized
targets which could be a nucleon or  nucleus. We also studied the nuclear dependence of the TMD parton distributions and
correlation functions and azimuthal asymmetries in reactions with unpolarized target nuclei.
Since future experiments at the Jefferson Laboratory for SIDIS with polarized nuclear targets are becoming possible \cite{HA},
it is of great interest to extend our previous study and provide some numerical estimates of the predicted nuclear
suppression of the azimuthal asymmetry for  polarized nuclear targets.
 In this Brief Report we provide an addendum to \cite{Song:2013sja} and
 extend the study of the nuclear dependence to reactions with polarized targets.
We also present numerical estimates of the nuclear dependence of the azimuthal asymmetry.

{\it Cross sections and azimuthal asymmetries}.
In the semi-inclusive process $e^-(l,s_l)+N(p,s)\to e^-(l^\prime)+q(k')+X$,
where variables in the brackets denote the momenta and polarizations,
the cross section and the azimuthal asymmetries can be obtained as functions of the
TMD parton distributions and/or correlation functions \cite{Song:2013sja,Bacchetta:2006tn},
\begin{align}
  \frac{d\sigma}{dx_Bdyd^2k_\perp}=&\frac{2\pi\alpha_{\rm em}^2e_q^2}{Q^2y}\big(W_{UU}+\lambda_lW_{LU}+s_\perp W_{UT}\nonumber\\
  &+\lambda W_{UL}+\lambda_l\lambda W_{LL}+\lambda_l s_\perp W_{LT}\big),
 \end{align}
where $W_{s_ls}$'s represent contributions in the different polarization cases
and are given by,
\begin{align}
  W_{UU}=&A(y)f_1-\frac{2x_B  |\vec  k_\perp|}{Q}B(y)f^\perp\cos\phi,\label{F-UU}\\
  W_{UT}=&\frac{ |\vec  k_\perp|}{M}A(y) f_{1T}^{\perp}\sin\left(\phi-\phi_s\right)\nonumber\\
  &- \frac{2x_BM}{Q} B(y) \Big[ f_T\sin\phi_s-\frac{k_\perp^2}{2M^2}f_T^\perp\sin(2\phi-\phi_s)\Big],\label{F-UT}\\
  W_{UL}=&-\frac{ 2x_B|\vec k_\perp|}{Q}B(y)f_L^\perp\sin\phi,\label{F-UL}\\
  W_{LU}=&-\frac{ 2x_B|\vec k_\perp|}{Q}D(y)g^\perp\sin\phi,\label{F-LU}\\
  W_{LL}=& C(y)g_{1L} - \frac{2x_B |\vec k_\perp|}{Q}D(y)g_L^\perp\cos\phi,\label{F-LL}\\
  W_{LT}=&\frac{|\vec k_\perp|}{M}C(y)g_{1T}^\perp \cos\left(\phi-\phi_s\right) \nonumber\\
    -& \frac{2x_BM}{Q} D(y) \Big[ g_T\cos\phi_s-\frac{k_\perp^2}{2M^2}g_T^\perp\cos\left(2\phi-\phi_s\right) \Big] .\label{F-LT}
\end{align}
where
\begin{eqnarray}
A(y)&=1+(1-y)^2, \;\;\; &B(y)=2(2-y)\sqrt{1-y}, \\
C(y)&=y(2-y), \;\;\;\;\;\;\;\;\;&D(y)=2y\sqrt{1-y},
\end{eqnarray}
$\cos\phi={\vec l_\perp\cdot \vec k_\perp}/{|\vec l_\perp||\vec k_\perp|}$,
$\sin\phi={(\vec l_\perp\times \vec k_\perp)\cdot \vec e_z}/{|\vec l_\perp||\vec k_\perp|}$,
$\cos\phi_s={\vec l_\perp\cdot \vec s_\perp}/{|\vec l_\perp||\vec s_\perp|}$,
$\sin\phi_s={(\vec l_\perp\times \vec s_\perp)\cdot\vec e_z}/{|\vec l_\perp||\vec s_\perp|}$;
 $y=p\cdot q/p\cdot l$, and $x_B$ is Bjorken-$x$  as fraction momentum carried by the struck quark.

The TMD parton distributions and/or correlation functions are defined via the decomposition of
the correlation matrix element that is given by,
\begin{align}
\hat\Phi^{(0)N}(x,k_\perp)&=\int \frac{p^+dy^- d^2y_\perp}{(2\pi)^3}  e^{ix p^+ y^- -i\vec{k}_{\perp}\cdot \vec{y}_{\perp}} 
\langle N|\bar{\psi}(0){\cal{L}}(0;y)\psi(y)|N\rangle\nonumber
\label{eq:Phi0def}\\
&=\left(\Phi^{(0)}_\alpha\gamma^\alpha-\tilde\Phi^{(0)}_\alpha\gamma_5\gamma^\alpha\right)/2+\ldots,
\end{align}
where $\Phi^{(0)}_\alpha$ and $\tilde\Phi^{(0)}_\alpha$ are decomposed as,
\begin{align}
\displaystyle\Phi^{(0)}_{\alpha}&=
\left(f_1-\varepsilon_\perp^{ks}f_{1T}^\perp\right)p_\alpha+f^\perp k_{\perp \alpha}
-f_TM\varepsilon_{\perp\alpha i}s_\perp^i\nonumber\\
&-\frac{f_T^\perp}{M}\Big(k_{\perp\alpha}k_{\perp\beta}-\frac{1}{2}k_\perp^2d_{\alpha\beta}\Big)\varepsilon_\perp^{\beta i}s_{\perp i}
-\lambda f_{L}^\perp\varepsilon_{\perp \alpha i}k_\perp^i+... \ , \label{phi0-lorentz}\\
 \tilde \Phi^{(0)}_{\alpha}&=-\Big(\lambda g_{1L}-\frac{k_\perp\cdot s_\perp}{M}g_{1T}^\perp\Big)p_\alpha
 + g^\perp\varepsilon_{\perp\alpha i}k_\perp^i-g_{T}Ms_{\perp \alpha}\nonumber\\
 &+\frac{g_T^\perp}{M}\Big(k_{\perp\alpha}k_{\perp\beta}-\frac{1}{2}k_\perp^2d_{\alpha\beta}\Big)s_{\perp}^{\beta}-\lambda g_{L}^\perp k_{\perp\alpha}+...
 \label{phi0t-lorentz}
\end{align}
There are two leading twist azimuthal asymmetries,
\begin{align}
\langle\sin(\phi-\phi_s)\rangle_{UT}&=s_\perp\frac{ |\vec  k_\perp|}{2M}\frac{f_{1T}^{\perp}(x,k_\perp)}{f_1(x,k_\perp)}, \label{aa:sinphi-phis} \\
\langle\cos(\phi-\phi_s)\rangle_{LT}&=\lambda_ls_\perp\frac{ |\vec  k_\perp|}{2M}\frac{C(y)}{A(y)}\frac{g^\perp_{1T}(x,k_\perp)}{f_1(x,k_\perp)}. \label{aa:cosphi-phis}
\end{align}
For $\langle \cos\phi\rangle$, we have,
\begin{align}
& \langle\cos\phi\rangle_{UU}=-\frac{|\vec  k_\perp|}{Q}\frac{B(y)}{A(y)}\frac{x_Bf^\perp(x_B,k_\perp)}{f_1(x_B,k_\perp)}, \label{aa:cosphiUU} \\
& \langle\cos\phi\rangle_{LL}=-\frac{ |\vec  k_\perp|}{Q}\nonumber\\
 &\phantom{XXX}\times \frac{B(y)x_Bf^\perp(x_B,k_\perp)+\lambda_l\lambda D(y)x_Bg^\perp_L(x_B,k_\perp)}
         {A(y)f_1(x_B,k_\perp)+\lambda_l\lambda C(y)g_{1L}(x_B,k_\perp)}, \label{aa:cosphiLL} \\
 &\langle\cos\phi\rangle_{LT}=\frac{ |\vec  k_\perp|}{2M} \nonumber\\
 &\times \frac{\lambda_l s_\perp C(y)g_{1T}^\perp(x_B,k_\perp)\cos\phi_s -\frac{2M}{Q}B(y)x_Bf^\perp(x_B,k_\perp)}
 {A(y)f_1(x_B,k_\perp)-\lambda_l s_\perp\frac{2M}{Q}x_Bg_T(x_B,k_\perp)\cos\phi_s}.  \label{aa:cosphiLT}
\end{align}
There are also twist-3 asymmetries $\langle \sin\phi\rangle$ for the $LU$ and $UL$ case given by,
 \begin{align}
\langle\sin\phi\rangle_{LU} &=
-\lambda_l\frac{|\vec k_\perp|}{Q}\frac{D(y)}{A(y)} \frac{x_Bg^\perp(x_B,k_\perp)}{f_1(x_B,k_\perp)},  \label{aa:sinphiLU} \\
\langle\sin\phi\rangle_{UL} &=
-\lambda \frac{|\vec k_\perp|}{Q}\frac{B(y)}{A(y)}\frac{x_Bf^\perp_L(x_B,k_\perp)}{f_1(x_B,k_\perp)}.  \label{aa:sinphiUL}
\end{align}
If we integrate over $\phi$, we obtain two transverse spin asymmetries at the twist-3 level,
\begin{align}
\langle\sin\phi_s\rangle_{UT} &=
-s_\perp\frac{M}{Q}\frac{B(y)}{A(y)}\frac{ x_Bf_T(x_B,k_\perp)}{f_1(x_B,k_\perp)},  \label{aa:sinphisUT} \\
\langle\cos\phi_s\rangle_{LT} &=
-\lambda_l s_\perp\frac{M}{Q}\frac{D(y)}{A(y)}\frac{x_Bg_T(x_B,k_\perp)}{f_1(x_B,k_\perp)}.  \label{aa:sinphisLT}
\end{align}
If we integrate over $|\vec k_\perp|$, we obtain, e.g.,
\begin{align}
  \langle\langle\sin\phi\rangle\rangle_{LU}=-\lambda_l\frac{B(y)}{A(y)}\frac{2\pi}{Q}\frac{\int \vec k_\perp^2 d|\vec k_\perp|  x_Bg^\perp(x_B,k_\perp)}{f_1(x_B)} .
\end{align}
These asymmetries all depend on the TMD parton distribution and correlation functions and apply to both nucleon and nuclear targets.

{\it Nuclear dependence}.
In SIDIS, the TMD quark distribution contains information of the interaction between the struck quark and the remnant of the target.
In a nucleus target, such interaction can occur between the struck quark and multiple nucleons inside the nucleus. Such multiple interaction
will lead to nuclear broadening of the TMD quark distribution.  Under the ``maximal two-gluon" or random-walk approximation,
the TMD quark distribution $\Phi^A_\alpha(x,k_\perp)$ in nucleus,
\begin{align}
\Phi^A_\alpha(x,k_\perp)\equiv & \int \frac{p^+dy^-d^2y_\perp}{(2\pi)^3}
e^{ixp^+y^- -i\vec  k_\perp\cdot \vec y_\perp} \nonumber\\
&\times\langle A \mid \bar\psi(0)\Gamma_\alpha{\cal L}(0;y)\psi(y)\mid A \rangle,
\label{form}
\end{align}
can be given by a convolution of the corresponding TMD quark distribution $\Phi^N_\alpha(x,k_\perp)$ in a nucleon
and a Gaussian broadening \cite{Liang:2008vz},
\begin{equation}
\Phi^A_\alpha(x,k_\perp)\approx\frac{A}{\pi \Delta_{2F}}
\int d^2\ell_\perp e^{-(\vec k_\perp -\vec\ell_\perp)^2/\Delta_{2F}}\Phi^N_\alpha(x,\ell_\perp),
\label{tmdgeneral}
\end{equation}
where $\Gamma_\alpha$ is any gamma matrix. The broadening width $\Delta_{2F}$, representing the total
transverse momentum broadening squared, is given by,
\begin{equation}
\Delta_{2F}=\int d\xi^-_N \hat q_F(\xi_N^-), 
\label{eq:Delta2F}
\end{equation}
where the quark transport parameter,
\begin{equation}
\hat q_F(\xi_N)
=\frac{2\pi^2\alpha_s}{N_c}\rho_N^A(\xi_N)[xf^N_g(x)]_{x=0},
\label{qhat1}
\end{equation}
is the effective transverse momentum broadening squared per unit distance for a fundamental quark which
is proportional to the nucleon number density  $\rho_N^A(\xi_N)$ and the gluon distribution $f^N_g(x)$ per nucleon.

The above relationship between the TMD quark distributions inside a nucleus and nucleon applies to both polarized and  unpolarized targets.
One can derive in particular~\cite{Gao:2010mj,Song:2013sja},
\begin{align}
  f_1^A(x,k_\perp)&\approx\frac{A}{\pi\Delta_{2F}}\int d^2\ell_\perp e^{-(\vec k_\perp -\vec\ell_\perp)^2/\Delta_{2F}}f_1^N(x,\ell_\perp),\\
  k_\perp^2f^{\perp A}(x,k_\perp)&\approx\frac{A}{\pi\Delta_{2F}}\int d^2\ell_\perp e^{-(\vec k_\perp -\vec\ell_\perp)^2/\Delta_{2F}}(k_\perp\cdot \ell_\perp)f^{\perp N}(x,\ell_\perp),\nonumber\\
  k_\perp^2g^{\perp A}(x,k_\perp)&\approx\frac{A}{\pi\Delta_{2F}}\int d^2\ell_\perp e^{-(\vec k_\perp -\vec\ell_\perp)^2/\Delta_{2F}}(k_\perp\cdot\ell_\perp)g^{\perp N}(x,\ell_\perp).\nonumber
\end{align}

To numerically estimate effects of the nuclear dependence of the TMD quark distribution,
we take a Gaussian ansatz ~\cite{Gao:2010mj,Song:2013sja} for TMD quark distributions in a nucleon,
\begin{align}
  &f_1^N(x,\ell_\perp)=\frac{1}{\pi\alpha}f_1^N(x)e^{-\vec \ell_\perp^2/\alpha},&\\
  &f^{\perp N}(x,\ell_\perp)=\frac{1}{\pi\beta}f^{\perp N}(x)e^{-\vec \ell_\perp^2/\beta},&\\
  &g^{\perp N}(x,\ell_\perp)=\frac{1}{\pi\gamma}g^{\perp N}(x)e^{-\vec \ell_\perp^2/\gamma}.
\end{align}
One can obtain the TMD quark distributions in a nucleus,
\begin{align}
  &f_1^A(x,k_\perp)\approx\frac{A}{\pi\alpha_A}f_1^N(x)e^{-\vec k_\perp^2/\alpha_A},&\\
  &f^{\perp A}(x,k_\perp)\approx\frac{A}{\pi\beta_A}\frac{\beta}{\beta_A}f^{\perp N}(x)e^{-\vec k_\perp^2/\beta_A},&\\
  &g^{\perp A}(x,k_\perp)\approx\frac{A}{\pi\gamma_A}\frac{\gamma}{\gamma_A}g^{\perp N}(x)e^{-\vec k_\perp^2/\gamma_A},\label{nuclearPDF:unpolarized}
\end{align}
where $\alpha_A=\alpha+\Delta_{2F}$, $\beta_A=\beta+\Delta_{2F}$, $\gamma_A=\gamma+\Delta_{2F}$.
We see that they all have a $k_\perp$-broadening with the same width $\Delta_{2F}$.
We also note that the twist-3 quark correlation functions $f^\perp(x,k_\perp)$ and $g^\perp(x,k_\perp)$
have an extra suppression factor $\beta/\beta_A$ or $\gamma/\gamma_A$.

The above approximation of TMD parton distributions and nuclear broadening can also be extended to SIDIS off polarized targets.
Here, we consider a nucleus with atomic number $A$ and spin $J_A$
and study the $A$-dependence of the spin related TMD parton distributions and correlation functions.
As a rough approximation, we assume that each nucleon has an equal polarization $J_A/A$ inside a nucleus.
In this case, the TMD quark distribution functions for a longitudinally polarized nucleus can be written similarly as,
\begin{align}
  g_{1L}^A(x,k_\perp)&\approx\frac{2J_A}{\pi\Delta_{2F}}\int d^2\ell_\perp e^{-(\vec k_\perp -\vec\ell_\perp)^2/\Delta_{2F}}g_{1L}^N(x,\ell_\perp),\\
  k_\perp^2f_L^{\perp A}(x,k_\perp)&\approx\frac{2J_A}{\pi\Delta_{2F}}\int d^2\ell_\perp
     e^{-(\vec k_\perp -\vec\ell_\perp)^2/\Delta_{2F}}(k_\perp\cdot\ell_\perp)f_L^{\perp N}(x,\ell_\perp),\nonumber\\
  k_\perp^2g_L^{\perp A}(x,k_\perp)&\approx\frac{2J_A}{\pi\Delta_{2F}}\int d^2\ell_\perp
     e^{-(\vec k_\perp -\vec\ell_\perp)^2/\Delta_{2F}}(k_\perp\cdot\ell_\perp)g_L^{\perp N}(x,\ell_\perp).\nonumber
\end{align}
These results are very similar as those for the unpolarized TMD distribution functions.
The only difference is that the overall multiplicative factor $A$ is now replaced by $2J_A$.
If we take the Gaussian ansatz for the TMD quark distribution functions with parameters
$\alpha^L$, $\beta^L$, and $\gamma^L$ for the longitudinally a polarized nucleon, we obtain,
\begin{align}
  &g_{1L}^A(x,k_\perp)\approx\frac{2J_A}{\pi\alpha^L_A}g_{1L}^N(x)e^{-\vec k_\perp^2/\alpha^L_A},\\
  &f_L^{\perp A}(x,k_\perp)\approx\frac{2J_A}{\pi\beta^L_A}\frac{\beta^L}{\beta^L_A}f_L^{\perp N}(x)e^{-\vec k_\perp^2/\beta^L_A},\\
  &g_L^{\perp A}(x,k_\perp)\approx\frac{2J_A}{\pi\gamma^L_A}\frac{\gamma^L}{\gamma^L_A}g_L^{\perp N}(x)e^{-\vec k_\perp^2/\gamma^L_A}.
\end{align}

Similarly, in the transversely polarized case, we obtain,
\begin{align}
& {\varepsilon_\perp^{ks}}\; f_{1T}^{\perp A}(x,k_\perp) \approx\frac{2J_A}{\pi\Delta_{2F}}\int d^2\ell_\perp e^{-(\vec k_\perp -\vec\ell_\perp)^2/\Delta_{2F}}\; \nonumber\\
&\phantom{XXXXXXX} \times {\varepsilon_\perp^{\ell s}}\; f_{1T}^{\perp N}(x,\ell_\perp),\\
 & (k_\perp\cdot s_\perp)g_{1T}^{\perp A}(x,k_\perp)  \approx\frac{2J_A}{\pi\Delta_{2F}}\int d^2\ell_\perp e^{-(\vec k_\perp -\vec\ell_\perp)^2/\Delta_{2F}} \nonumber\\
  &\phantom{XXXXXXX} \times (\ell_\perp\cdot  s_\perp)g_{1T}^{\perp N}(x,\ell_\perp),\\
 &{\varepsilon_\perp^{ks}}(k_\perp\cdot s_\perp) f_T^A(x,k_\perp)  \approx\frac{2J_A}{\pi\Delta_{2F}}\int d^2\ell_\perp e^{-(\vec k_\perp -\vec\ell_\perp)^2/\Delta_{2F}} \nonumber\\
 &\phantom{XXX}\times \Big\{{\varepsilon_\perp^{ks}}(k_\perp\cdot s_\perp)  f_T^N(x,\ell_\perp)
 -\Big[(k_\perp\cdot s_\perp)(k_\perp\cdot \ell_\perp)\varepsilon_\perp^{\ell s}\nonumber\\
  &\phantom{XXX} -(k_\perp\cdot s_\perp)\varepsilon_\perp^{ks}\frac{\ell_\perp^2}{2}-(\ell_\perp\cdot s_\perp)\varepsilon_\perp^{\ell s}\frac{k_\perp^2}{2}\Big]\frac{f_T^{\perp N}(x,\ell_\perp)}{M^2}\Big\},\\
&  {\varepsilon_\perp^{ks}}(k_\perp\cdot s_\perp) g_T^{A}(x,k_\perp)\approx\frac{2J_A}{\pi\Delta_{2F}}\int d^2\ell_\perp e^{-(\vec k_\perp -\vec\ell_\perp)^2/\Delta_{2F}} \nonumber\\
&\phantom{XXX}\times \Big\{{\varepsilon_\perp^{ks}}(k_\perp\cdot s_\perp) g_T^{ N}(x,\ell_\perp)+\Big[\varepsilon_\perp^{ks}(k_\perp\cdot \ell_\perp)(\ell_\perp\cdot s_\perp)\nonumber\\
&\phantom{XXX}-(k_\perp\cdot s_\perp)\varepsilon_\perp^{ks}\frac{\ell_\perp^2}{2}-(\ell_\perp\cdot s_\perp)\varepsilon_\perp^{\ell s}\frac{k_\perp^2}{2}\Big]\frac{g_T^{\perp N}(x,\ell_\perp)}{M^2}\Big\},\\
 & {\varepsilon_\perp^{ks}}(k_\perp\cdot s_\perp) f_T^{\perp A}(x,k_\perp)\approx\frac{2J_A}{\pi\Delta_{2F}}\int d^2\ell_\perp e^{-(\vec k_\perp -\vec\ell_\perp)^2/\Delta_{2F}} \nonumber\\
 & \phantom{XXXXXXX}\times {\varepsilon_\perp^{\ell s}}(\ell_\perp\cdot  s_\perp) f_T^{\perp N}(x,\ell_\perp),\\
&  {\varepsilon_\perp^{ks}}(k_\perp\cdot s_\perp) g_T^{\perp A}(x,k_\perp)\approx\frac{2J_A}{\pi\Delta_{2F}}\int d^2\ell_\perp e^{-(\vec k_\perp -\vec\ell_\perp)^2/\Delta_{2F}}\nonumber\\
& \phantom{XXXXXXX}\times {\varepsilon_\perp^{\ell s}}(\ell_\perp\cdot s_\perp ) g_T^{\perp N}(x,\ell_\perp).
\end{align}
Taking the Gaussian ansatz, we obtain,
\begin{align}
  &f_{1T}^{\perp A}(x, k_\perp)\approx\frac{2J_A}{\pi\alpha^T_A}\frac{\alpha^T}{\alpha_A^T}f_{1T}^{\perp N}(x)e^{-\vec  k_\perp^2/\alpha^T_A},&\\
  &g_{1T}^{\perp A}(x, k_\perp)\approx\frac{2J_A}{\pi\tilde \alpha^T_A}\frac{\tilde \alpha^T}{\tilde \alpha_A^T}g_{1T}^{\perp N}(x)e^{-\vec  k_\perp^2/\tilde \alpha^T_A},&\\
  &f_T^A(x, k_\perp)\approx\frac{2J_A}{\pi\beta^T_A}f_T^N(x)e^{-\vec  k_\perp^2/\beta^T_A},&\\
  &g_T^A(x, k_\perp)\approx\frac{2J_A}{\pi\tilde \beta^T_A}g_T^N(x)e^{-\vec  k_\perp^2/\tilde \beta^T_A},&\label{f-T-A}\\
  &f_T^{\perp A}(x, k_\perp)\approx\frac{2J_A}{\pi\gamma^T_A}\left(\frac{\gamma^T}{\gamma_A^T}\right)^2f_T^{\perp N}(x)e^{-\vec  k_\perp^2/\gamma^T_A},&\label{f-T-perp-A}\\
  &g_T^{\perp A}(x, k_\perp)\approx\frac{2J_A}{\pi\tilde \gamma^T_A}\left(\frac{\tilde\gamma^T}{\tilde\gamma_A^T}\right)^2g_T^{\perp N}(x)e^{-\vec  k_\perp^2/\tilde \gamma^T_A}, &
\end{align}
where the superscript $T$ for $\alpha$, $\beta$ and $\gamma$ denotes transversely polarized nucleon and
all the widths $\alpha$, $\beta$ and $\gamma$ for the nucleus are equal to the corresponding ones for
the nucleon plus the broadening width $\Delta_{2F}$.
We note that there are extra suppression factors such as $\alpha^T/\alpha^T_A$.
Generally we can observed that the power of the extra suppression factor concide with the power of $k_\perp$ in the tensor decomposition formulae Eq.(\ref{phi0-lorentz})(\ref{phi0t-lorentz}).


Using these results for TMD parton correlations,
we obtain the nuclear dependences of azimuthal asymmetries.
For reactions with unpolarized targets, up to twist-3, we have \cite{Song:2013sja},
\begin{align}
\frac{\langle\cos\phi\rangle_{UU}^{eA}}{\langle\cos\phi\rangle_{UU}^{eN}}\approx&\frac{\alpha_A}{\alpha}\Big(\frac{\beta}{\beta_A}\Big)^2
  e^{\big(\frac{1}{\alpha_A}-\frac{1}{\alpha}-\frac{1}{\beta_A}+\frac{1}{\beta}\big)\vec k_\perp^2}, \\
  \frac{\langle\sin\phi\rangle_{LU}^{eA}}{\langle\sin\phi\rangle_{LU}^{eN}}\approx&\frac{\alpha_A}{\alpha}\Big(\frac{\gamma}{\gamma_A}\Big)^2
  e^{\big(\frac{1}{\alpha_A}-\frac{1}{\alpha}-\frac{1}{\gamma_A}+\frac{1}{\gamma}\big)\vec k_\perp^2}.
\end{align}

If $\alpha=\beta=\gamma$, they reduce to the $k_\perp$-independent results,
\begin{align}
  \frac{\langle\cos\phi\rangle_{UU}^{eA}}{\langle\cos\phi\rangle_{UU}^{eN}}\approx
  \frac{\langle\sin\phi\rangle_{LU}^{eA}}{\langle\sin\phi\rangle_{LU}^{eN}}\approx\frac{\alpha}{\alpha+\Delta_{2F}}.
\end{align}
After integrating over $|\vec k_\perp|$, we have,
\begin{align}
 \frac{\langle\langle\cos\phi\rangle\rangle_{UU}^{eA}}{\langle\langle\cos\phi\rangle\rangle_{UU}^{eN}}\approx
 \frac{\langle\langle\sin\phi\rangle\rangle_{LU}^{eA}}{\langle\langle\sin\phi\rangle\rangle_{LU}^{eN}}\approx
 &\sqrt{\frac{\alpha}{\alpha+\Delta_{2F}}}.
\end{align}

For reactions with polarized targets, the two leading twist azimuthal asymmetries in
Eqs.~(\ref{aa:sinphi-phis}) and (\ref{aa:cosphi-phis}) are given by,
\begin{align}
\frac{\langle\sin(\phi-\phi_s)\rangle_{UT}^{eA}}{\langle\sin(\phi-\phi_s)\rangle_{UT}^{eN}}&
    \approx\frac{2J_A}{A}\frac{\alpha_A}{\alpha}\Big(\frac{\alpha^T}{\alpha^T_A}\Big)^2
    e^{\big(\frac{1}{\alpha_A}-\frac{1}{\alpha}-\frac{1}{\alpha_A^T}+\frac{1}{\alpha^T}\big)\vec k_\perp^2},\\
\frac{\langle\cos(\phi-\phi_s)\rangle_{UT}^{eA}}{\langle\cos(\phi-\phi_s)\rangle_{UT}^{eN}}&
    \approx\frac{2J_A}{A}\frac{\alpha_A}{\alpha}\Big(\frac{\tilde\alpha^T}{\tilde\alpha^T_A}\Big)^2
    e^{\big(\frac{1}{\alpha_A}-\frac{1}{\alpha}-\frac{1}{\tilde\alpha_A^T}+\frac{1}{\tilde\alpha^T}\big)\vec k_\perp^2} .
\end{align}
For the twist-3 azimuthal asymmetries, we obtain,
\begin{align}
\frac{\langle\sin\phi\rangle_{UL}^{eA}}{\langle\sin\phi\rangle_{UL}^{eN}}&
 \approx\frac{2J_A}{A}\frac{\alpha_A}{\alpha}\Big(\frac{\beta^L}{\beta^L_A}\Big)^2
 e^{\big(\frac{1}{\alpha_A}-\frac{1}{\alpha}-\frac{1}{\beta^L_A}+\frac{1}{\beta^L}\big)\vec k_\perp^2},\\
\frac{\langle\sin\phi_s\rangle_{UT}^{eA}}{\langle\sin\phi_s\rangle_{UT}^{eN}}&
 \approx\frac{2J_A}{A}\frac{\alpha_A}{\alpha}\Big(\frac{\beta^T}{\beta^T_A}\Big)
 e^{\big(\frac{1}{\alpha_A}-\frac{1}{\alpha}-\frac{1}{\beta^T_A}+\frac{1}{\beta^T}\big)\vec k_\perp^2}, \\
 \frac{\langle\cos\phi_s\rangle_{LT}^{eA}}{\langle\cos\phi_s\rangle_{LT}^{eN}}&
 \approx\frac{2J_A}{A}\frac{\alpha_A}{\alpha}\Big(\frac{\tilde\beta^T}{\tilde\beta^T_A}\Big)
 e^{\big(\frac{1}{\alpha_A}-\frac{1}{\alpha}-\frac{1}{\tilde\beta^T_A}+\frac{1}{\tilde\beta^T}\big)\vec k_\perp^2}.
\end{align}
If all the widths for the transverse momentum dependence are taken as the same,
these ratios become equal and $k_\perp$-independent,
\begin{align}
&\frac{\langle\sin(\phi-\phi_s)\rangle_{UT}^{eA}}{\langle\sin(\phi-\phi_s)\rangle_{UT}^{eN}}
=\frac{\langle\cos(\phi-\phi_s)\rangle_{UT}^{eA}}{\langle\cos(\phi-\phi_s)\rangle_{UT}^{eN}}
=\frac{\langle\sin\phi\rangle_{UL}^{eA}}{\langle\sin\phi\rangle_{UL}^{eN}}\approx \frac{2J_A}{A}\frac{\alpha}{\alpha_A},\\
&\frac{\langle\sin\phi_s\rangle_{UT}^{eA}}{\langle\sin\phi_s\rangle_{UT}^{eN}}
=\frac{\langle\cos\phi_s\rangle_{LT}^{eA}}{\langle\cos\phi_s\rangle_{LT}^{eN}}
    \approx\frac{2J_A}{A}.
\end{align}
We see that the asymmetries in the case with a polarized nucleus target are in general much more
suppressed by the dilution factor $2J_A/A$.

The situations for $\langle\cos\phi_s\rangle_{LL}$ and $\langle\cos\phi_s\rangle_{LT}$
are more complicated because of the competition between the two terms in the numerators
and the denominators. No such simple results can be obtained.

{\it  Numerical estimates}.
As discussed in earlier, if the widths of the nucleon TMD parton distributions are taken as the same, i.e.
$\alpha=\beta=\gamma$ and they are the same for longitudinally or transversely polarized nucleon,
all the asymmetries in the unpolarized case are suppressed by a factor $\alpha/(\alpha+\Delta_{2F})$
and those in reactions with polarized targets are further suppressed by the dilution factor $2J_A/A$.
We see that the most important factor describing the nuclear dependence
is $\alpha/(\alpha+\Delta_{2F})$ that we denote by a suppression factor $f_s$ in the following.

The suppression factor $f_s$ is determined by two parameters.
The constant $\alpha$ is the width of the Gaussian ansatz for the quark's transverse momentum distribution in a nucleon.
Approximately, it is taken as $\alpha=0.2 \sim 0.3$ GeV$^2$ (see e.g \cite{Anselmino:2013rya}).

The other parameter is the nuclear broadening width $\Delta_{2F}$ given by Eq.~(\ref{eq:Delta2F}).
We assume a hard-sphere nuclear distribution for $\rho_N^A(\xi_N,b)$ with a normalization $\int d\xi_N d^2b \rho_N^A(\xi_Nb)=A$.
The quark transport parameter is proportional to the nuclear density
\begin{equation}
\hat q_F(\xi, b)=\hat q_0 {\rho_N^A(\xi,b)}/{\rho_N^A(0,0)},
\end{equation}
where $\hat q_0$ is the quark transport parameter at the center of a nucleus. Since the probability of $\gamma^*N$
interaction inside a nucleus is assumed to be proportional to $\rho_N^A(\xi,b)$, the averaged transverse momentum broadening
in DIS is then
\begin{eqnarray}
\Delta_{2F}&=&\frac{1}{A} \int_{-\infty}^\infty d\xi_N d^2b  \int_{\xi_N}^\infty d\xi \hat q_F(\xi, b) \rho_N^A(\xi_N,b) \nonumber \\
&=& {3\sqrt{2}}\hat q_0r_0A^{1/3}/4,
\end{eqnarray}
where $R_A=r_0A^{1/3}$ is the nuclear radius with $r_0\approx 1.12$ fm and the $\sqrt{2}$ comes from the $\xi^-\to \xi_z$ transformation.
The nuclear suppression factor for the azimuthal asymmetry is then.
\begin{align}
  f_s \approx (1+{3\sqrt{2}}\hat q r_0A^{1/3}/{4\alpha})^{-1}.
\end{align}
 From analyses of jet suppression of leading hadrons in SIDIS of large nuclei due to multiple scattering,
 the quark transport parameter at the center of a large nucleus has been determined to be
  $\hat q_0\approx 0.024\pm 0.008$ GeV$^2$/fm  \cite{Wang:2009qb,Chang:2014fba}.
  Assuming $\alpha=0.25$~GeV$^2$, we have
 \begin{align}
  f_s\approx (1+0.114A^{1/3})^{-1}.
\end{align}
In Fig. 1, we plot $f_s$ as a function of $A$.
Recent measurements on transverse momentum broadening in SIDIS have been carried out by HERMES\cite{Airapetian:2009jy,Airapetian:2011jp}. 
The transverse momentum of the final hadron can be related to that of the final parton approximately, $p_{h\perp}\approx zk_\perp+p_{f\perp}$, where $p_{f\perp}$ is the transvsre 
momentum of hadron obtained in the fragmentation. 
By neglecting the nuclear modifications on 
the $z$-dependence and the $p_{f\perp}$ part, 
we obtain a simple rough estimation of $\Delta \langle p_{h\perp}^2\rangle$ as $\Delta \langle p_{h\perp}^2\rangle\approx z^2\Delta_{2F}$. 
We compare this rough estimation with data\cite{Airapetian:2009jy,Airapetian:2011jp}
in Fig.~1 and we see that the qualitative tendency 
is reproduced.\cite{ff}

\begin{figure}[ht!]
  \centering
  \epsfig{file=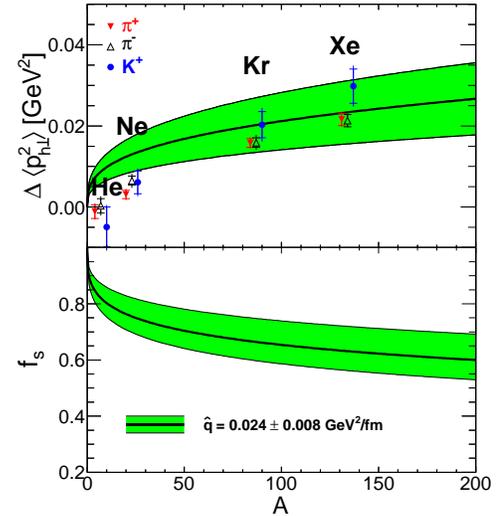,width=0.85\linewidth,clip=}
  \caption{The nuclear suppression factor for azimuthal asymmetry $f_s$ as a function of $A$ (lower part) 
  and a comparison the transverse momentum broadening 
  with data\cite{Airapetian:2009jy,Airapetian:2011jp} (upper part). }
\end{figure}

{\it Summary}.
In summary, we have extended the study of nuclear dependence of azimuthal asymmetries
in \cite{Song:2013sja} to SIDIS off polarized targets.  The results show a further suppression factor $2J_A/A$
as compared to the unpolarized SIDIS. We also present a simple numerical estimate of the
broadening width $\Delta_{2F}$ and the nuclear suppression factor for azimuthal asymmetries
that might be helpful in guiding future experiments.

{\it Acknowledgements}. We thank Harut Avakian for communications.
This work was supported in part by the National Natural Science Foundation of China
(project 11035003, 11105137  and  11221504),  
the Major State Basic Research Development Program in China (No. 2014CB845404), 
the Ministry of Science and Technology of China under grant No. 2014DFG02050,
Office of Energy Research, Office of High Energy and Nuclear Physics,
Division of Nuclear Physics, of the U.S. Department of Energy under Contract No.
DE-AC02-05CH11231, and by CCNU-QLPL Innovation Fund (QLPL2011P01) .


\begin{thebibliography}{}


\bibitem{0005044}
  X.~-F.~Guo and X.~-N.~Wang,
  Phys.\ Rev.\ Lett.\  {\bf 85}, 3591 (2000).

\bibitem{0102230}
  X.~-N.~Wang and X.~-F.~Guo,
  Nucl.\ Phys.\ A {\bf 696}, 788 (2001).

\bibitem{9804234}
  X.~-F.~Guo,
  Phys.\ Rev.\ D {\bf 58}, 114033 (1998).

\bibitem{Kang:2013raa}
  Z.~-B.~Kang, E.~Wang, X.~-N.~Wang and H.~Xing,
  arXiv:1310.6759 [hep-ph].


\bibitem{Gao:2010mj}
  J.~-H.~Gao, Z.~-T.~Liang and X.~-N.~Wang,
  Phys.\ Rev.\ C {\bf 81}, 065211 (2010).
  %


\bibitem{Song:2013sja} 
  Y.~-K.~Song, J.~-h.~Gao, Z.~-T.~Liang and X.~-N.~Wang,
  Phys.\ Rev.\ D {\bf 89}, 014005 (2014).

\bibitem{HA}
Harut Avakian, private communications (2013).

\bibitem{Bacchetta:2006tn}
  A.~Bacchetta, M.~Diehl, K.~Goeke, A.~Metz, P.~J.~Mulders and M.~Schlegel,
  JHEP {\bf 0702}, 093 (2007).


\bibitem{Liang:2008vz}
  Z.~-T.~Liang, X.~-N.~Wang and J.~Zhou,
  Phys.\ Rev.\ D {\bf 77}, 125010 (2008).



\bibitem{Anselmino:2013rya}
  M.~Anselmino, M.~Boglione, U.~D'Alesio, S.~Melis, F.~Murgia and A.~Prokudin,
  Phys.\ Rev.\ D {\bf 88}, 054023 (2013).

\bibitem{Wang:2009qb}
  W.~-T.~Deng and X.~-N.~Wang,
  Phys.\ Rev.\ C {\bf 81}, 024902 (2010).

\bibitem{Chang:2014fba}
  N.~-B.~Chang, W.~-T.~Deng and X.~-N.~Wang,
  arXiv:1401.5109 [nucl-th].

\bibitem{Airapetian:2009jy} 
  A.~Airapetian {\it et al.}  [HERMES Collaboration],
  Phys.\ Lett.\ B {\bf 684}, 114 (2010)
  [arXiv:0906.2478 [hep-ex]].


\bibitem{Airapetian:2011jp} 
  A.~Airapetian {\it et al.}  [HERMES Collaboration],
  Eur.\ Phys.\ J.\ A {\bf 47}, 113 (2011)
  [arXiv:1107.3496 [hep-ex]].

\bibitem{ff} The $z$-dependence of $\Delta\langle p_{h\perp}^2\rangle$ 
should receive much influence from the nuclear modification of the $z$-dependence 
part of the fragmentation function and also the formation time of hadron that is estimated 
proportional to $z(1-z)$. We do not go to the details here. 



\end{thebibliography}
\end{document}